\documentclass[pdflatex,sn-mathphys-ay]{sn-jnl}

\usepackage{graphicx}%
\usepackage{multirow}%
\usepackage{amsmath,amssymb,amsfonts}%
\usepackage{amsthm}%
\usepackage{mathrsfs}%
\usepackage[title]{appendix}%
\usepackage{xcolor}%
\usepackage{textcomp}%
\usepackage{manyfoot}%
\usepackage{booktabs}%
\usepackage{algorithm}%
\usepackage{algorithmicx}%
\usepackage{algpseudocode}%
\usepackage{listings}%

\theoremstyle{thmstyleone}%
\theoremstyle{thmstyletwo}%

\theoremstyle{thmstylethree}%

\begin{document}

\title[On the independence problem of Newton’s first law]{On the independence problem of Newton’s first law}


\author[1]{\fnm{Ido} \sur{Yavetz}}\email{yavetz@tauex.tau.ac.il}

\author*[1,2]{\fnm{Ehud} \sur{Aharoni}}\email{aehud@il.ibm.com}

\affil[1]{\orgdiv{The Cohn Institute for the History and Philosophy of Science and Ideas}, \orgname{Tel Aviv University}, \orgaddress{\street{P.O. Box 39040},\city{Tel Aviv}, \postcode{6997801}, \country{Israel}}}

\affil[2]{\orgdiv{IBM Research - Israel}, \orgname{IBM Research}, \orgaddress{\street{Haifa University Campus},\city{Mount Carmel Haifa}, \postcode{3498825}, \country{Israel}}}

\abstract{Newton’s laws of motion pose an apparent problem, sometimes referred to as “the independence problem”: the first law seems to be a simple consequence of the second law, raising the question of why it was included as a separate law. Numerous answers to this question have been proposed in the literature.

The main contribution of this paper is a novel answer which we call “the formal explanation.” Unlike previous accounts it relies on mathematical formalism and  argues that the definitions of Euclidean geometry necessitate the inclusion of the first law. We provide evidence in support of this claim. A second contribution is a comprehensive review of previously suggested explanations, which so far have often been treated in a fragmented manner,
and a discussion of the plausibility of the various answers.

}




\maketitle

\section{Introduction}
Newton’s laws of motion contain within them a puzzle: the first law seems redundant. This law, reformulated in modern English, states that “Every body not subject to the action of forces continues in its state of rest or uniform motion in a straight line"~\citep[35]{Ellis1965}. 
The modern formulation of the second law is the equation $F=ma$: the force equals the product of mass and acceleration. Substituting $F=0$ into this equation yields $a=0$, hence the second law implies that in the absence of forces a body has zero acceleration. This statement is equivalent to the first law and therefore the first law is seemingly an implication of the second.

These two laws (together with the third one, the law of action and reaction) are listed in Newton’s \emph{Principia} under the title “Axioms, or the laws of motion.” The question is therefore, why did Newton include as an axiom a statement that appears to be a simple consequence of another axiom?

This problem, sometimes referred to as “the independence problem”~\citep{Hoek2023}, has received some attention in the literature. Thomas Kuhn writes about it:

“Newton’s first law is a consequence of his second, and Newton’s reason for stating them separately has long been a puzzle.”~\citep{Kuhn2000}

Numerous authors have attempted to resolve this puzzle, offering a variety of conjectures. Some suggest, for instance, that the first law conveys a deeper meaning beyond its literal wording, thereby showing it is not redundant. Others maintain that the first law is indeed redundant, but was included for rhetorical or pedagogical purposes. We discuss these and many other suggested solutions at length below.

Furthermore, we propose a new explanation, which we call \emph{the formal explanation}, according to which the inclusion of the first law arises from a technical requirement of Euclidean geometry. We argue that this consideration was likely the primary reason for the law’s formulation, and provide evidence in support of this claim.
We further propose a second answer, which we term \emph{the logical explanation}, stating that the first law's importance is due to capturing a more basic and general property of nature. 
Similar views are already known in the literature, but we present a new variant and argue that it may have served as a supporting reason as well.

Note that this paper focuses on the historical question: Why did \emph{Newton} formulate the first law separately? A different question is whether the first law is required as an independent postulate in modern formulations of classical mechanics, given modern mathematical tools and concepts such as inertial frames that were unavailable to Newton. For example, ~\cite{Earman1973} provide such an account. This philosophical question, however, lies outside the scope of the present paper. Here we focus solely on Newton’s reasons for distinguishing the first law from the second.

The rest of the paper is organized as follows. Section~\ref{sec:review} reviews the relevant literature. Section~\ref{sec:novel} presents our novel account. Section~\ref{sec:discussion} provides additional arguments in favor of our approach and discusses the plausibility of competing explanations of the independence problem. Section~\ref{sec:conclusion} concludes.

\section{Literature review}
\label{sec:review}
In this section, we provide an overview of the different solutions to the independence problem offered in the literature. 

\subsection{Conceptual precedence}
We begin with what is perhaps the most prevalent type of solution: the view that the first law is a conceptual prerequisite for the second law. Without it, the second law becomes incomplete or entirely unintelligible. In the words of Tim Maudlin:

“The Second Law is conceptually parasitic on the First Law"~\citep[19]{Maudlin2012}

\subsubsection{The inertial frame explanation}
Many physics textbooks state that the role of the first law is to define what an inertial frame of reference is, to identify it, or to postulate that such a frame exists:\footnote{In addition to the three books cited here, see more books cited in~\citep{Galili2003,Earman1973}.}

“The real assertion [of the first law] is, rather: There exists a coordinate system (or coordinate systems) with respect to which all bodies not subjected to forces are unaccelerated”.~\citep[8]{Bergmann1976}

“Another thing [the first law] does is give a definition of an inertial frame.”~\citep[52]{Morin2004}

“Newton’s first law of motion is the assertion that inertial systems exist”~\citep[51]{Kleppner2013}

Some of these books are vague whether this was the original reason the first law was introduced or whether it is a new modern interpretation of the laws of motion. This vagueness may have contributed to a confusion over this matter, causing a quite widespread misconception that Newton offered this explanation himself, as can be found in some books and web-based publications.

Researchers that address the independence problem clearly separate the discussion between possible modern explanations and Newton’s original reasons. The term “inertial frame” did not exist in Newton’s time; instead, he used a different concept --- “Absolute space.” This concept, as other concepts used by the laws of motion, is explained in the \emph{Principia} in a separate “Definitions” section preceding that of the laws. Hence most scholarly works on this subject reject the plausibility of the inertial frame explanation, arguing that the \emph{Principia}’s text offers no support for it.

Zinkernagel, for example, accepts the inertial frame explanation as a modern explanation, but rejects it as a possible reason for Newton:

“Now, one way to understand Newton’s first law in a non-redundant way is precisely to construct it as an existence claim . . .  On this account, Newton’s first law implies that inertial frames . . .  exist. Newton himself did not need the first law to secure the existence of an inertial frame as he independently (of the laws) assumed the existence of absolute space.”~\citep[5]{Zinkernagel2010}

Earman and Friedman agree:

“In the \emph{Principia}, the statement of Law I,  . . . immediately follows the Scholium on Absolute Space. Thus, it is clear that Law I is to be taken to refer to motion with respect to Absolute Space  . . . In any case, Newton did not intend nor did he have any need to use the First Law as a definition of inertial frame; because of the assumed existence of Absolute Space, Law I made good sense as it stood.”~\citep[338]{Earman1973}


The association between the first law and the definition of inertial frames began in the late 19\textsuperscript{th} century, with early definitions of this concept~\citep{DiSalle2020}. 
As noted above, the proposed modern interpretations that justify the independence of the first law are outside the scope of this paper. We focus solely on Newton’s reasons for stating the first law separately.\footnote{We did find one scholarly work stating that the inertial frame explanation may have been Newton’s reason for stating the first law separately~\citep{OSullivan1980}, though they do not address the question of how the above-mentioned structure of the \emph{Principia}’s text can be reconciled with such a view.}

\subsubsection{Definition of natural motion}

Some authors argue that the first law’s role is to define natural motion, the motion that occurs in the absence of external causes. 
On this view, the second law cannot be fully understood without first specifying what counts as natural motion, and hence the first law  is required as a separate law.
Pourciau and Maudlin express such a view:

"in its compound interpretation, the second law predicts the force by measuring the deflection from L, the place the body would have been had there been no force, yet only through the first law do we know where to find the place L, and so without a separate and prior statement of the first law, the second law has no clear meaning!"~\citep[189]{pourciau2006}
    
“without the definition of a ‘state of motion’ in the First Law, we could make no sense of ‘change in motion’ in the Second. According to Aristotle and to Galileo, certain uniform circular motions are constant motions: they display no change of motion, and so they call for no special explanation. But according to Newton, a body in uniform circular motion is constantly changing its state of motion, and so it must be subject to some external force.”~\citep[19]{Maudlin2012}

According to this view, the second law specifies how a body's motion changes under the influence of forces,
but it does not specify what counts as constant, unaltered motion. It is the first law that defines unaltered motion as motion in a straight line at constant speed.

Pourciau~\citep{pourciau2006} provides a detailed account in support of this view. His main focus is actually on a separate issue: whether the force mentioned in the second law is an impulse force or a continuously acting force. He advocates the "compound interpretation", according to which the second law covers both kinds of forces, and explains how this interpretation also resolves the independence problem.

Pourciau argues that the term "Change in motion" used by the second law does not refer to the term "Quantity of motion" defined by Newton in the Definitions as the combination of quantity of matter and velocity, but rather to the notion of motion introduced in the scholium to the Definitions:

"Absolute motion is the change of position of a body from one absolute place to another"~\citep{newton1999principia}.

Thus, the second law does not specify how forces change velocity, but how they change the position a body will be in compared to the position it would have had in the absence of force. The first law's role is to specify this position. One argument in support of this view is an alternative formulation of the second law found in a manuscript in which Newton considered revisions to the second edition:

"All new motion by which the state of a body is changed is proportional to the motive force impressed, and occurs from the place which the body would otherwise occupy towards the goal at which the impressed force aims." ~\citep[539]{math_papers_of_newton}

\subsubsection{Other concepts}

\cite{Zinkernagel2010} claims the first law’s role is to define absolute time. Here too he may suggest this explanation as a modern interpretation and not as the original reason for this law. Similarly, \cite{Arons1997} states the first law provides a qualitative operational definition of a force, again without attributing this explanation directly to Newton.

Earman and Friedman provide a similar explanation, but they do refer to Newton directly:

“Secondly, Law I is psychologically if not logically the source of Law II for Newton since he introduces forces as the cause of deviation from inertial motion.”~\citep[346]{Earman1973}

\cite{Ellis1965} claims that “force” is not an object that exists in reality, but a convention only, arising from us choosing to treat uniform motion as the natural motion. The first law’s role is to define what a natural motion is, and thus also what a force is. This view is similar to that of Maudlin and Pourciau quoted above, except it further attributes the first law with a definition of force.

Another similar view is by \cite{Koyre1965}, who also claims that it is the first law that asserts the meaning of a state of motion. His view is somewhat different than the view of Maudlin and Pourciau detailed in the previous subsection.  Maudlin and Pourciau considered the first law as defining what counts as unaltered motion. Koyré assigns the first law with a more basic task: an assertion that motion can be considered to be a state just like rest is a state. 
This in the context of the previous belief since Aristotle that motion is a process of change and not a state, and therefore requires a cause.

"by using this expression Newton implies or asserts that motion is not, as had been believed for about two thousand years - since Aritsotle - a process of change, in contradistinction to rest, which is truely a status, but is a also a state, that is, something that no more implies change than does rest."~\citep[66]{Koyre1965}

According to this view, before stating a law that details how a motion changes, we must first establish that motion itself is not a process of change.

\subsection{Accepting the redundancy}
\label{sec:review-accept}
A second general strategy to solve the independence problem is to accept that the first law is in fact redundant. On this view, it was included not out of logical necessity, but for other reasons.

\subsubsection{Pedagogical reasons}
\label{sec:review-ped}
Kuhn suggests that the reason for the first law is pedagogical. The second law uses the terms mass and force, which makes it more difficult to understand  (Newton’s original formulation does not use the term mass, but rather motion, and through Definition 2 we understand it means what we call today momentum). Especially so since Newton used these terms with different meanings than his predecessors. The first law was therefore added to help the readers take their first steps in understanding the new theory.

“The answer may well lie in pedagogical strategy. If Newton had permitted the second law to subsume the first, his readers would have had to sort out his use of ‘force’ and of ‘mass’ together, an intrinsically difficult task further complicated by the fact that the terms had previously been different not only in their individual use but in their interrelation. Separating them to the extent possible displayed the nature of the required changes more clearly.”~\citep[68]{Kuhn2000} 

\subsubsection{Rhetorical reasons}
\label{sec:review-rhet}
Other authors have suggested that the reason is rhetorical. They acknowledge that the first law is not independent and argue that by stating it separately Newton sought to emphasize the difference between his new theory and its predecessors, especially the Aristotelian theory:

“Hence, we can undercut the search for an independent message hidden in the first law by viewing it as inaugurating a new ideal of natural order. In other word, the real role of the first law is as a slap in the face of common sense and Aristotle. It is intended to jolt us into a new way of looking at nature.”~\citep[10]{Sorensen1992}

Cohen makes a similar statement:

“First, in Newton’s day – as during many preceding centuries – the common belief was that all motion requires a mover, a moving force. The very statement of this law as an axiom was a radical step, a declaration of an important new principle of motion, too important to be a special case of another law.”~\citep[69]{Cohen2002}\footnote{Additional authors expressing a view similar to Cohen’s are~\cite{Galili2003}. By studying various translations of the \emph{Principia}, they conclude that the first law not only states that in the absence of a force an object preserves its state, but that it also implies that forces cause a sequence of state changes.
They use this extended meaning of the first law to support the view that the first law is more general than the second (similarly to our logical explanation described in Section~\ref{sec:log}). However, like Cohen, they argue that the importance of the first law lies in expressing a radical conceptual change in the understanding of motion.}

\subsubsection{Historical influence}
\label{sec:review-hist}
Another solution to the independence problem is that Newton’s choice to formulate two separate laws is due to influence from his predecessors, specifically Descartes, Galileo, and Huygens~\citep{newton1999principia,Cohen2002}.

Descartes is identified as the source of the first law and the notion that a uniform motion constitutes a state. Huygens, identified as the most important influence, formulated two hypotheses similar to Newton’s first two laws. The first hypothesis stated that a body free of air resistance and gravity would move with uniform velocity in a straight line. The second hypothesis described the motion of a body under the influence of gravity. 

In a statement in the \emph{Principia}, Newton claims that Galileo knew of the two laws of motion and derived his results from them. This may be indicative of influence by previous works, though Newton’s knowledge of Galileo may have come from secondary sources.

\subsection{Narrowing the scope of the second law}
\label{sec:review-narrow}
Faced with the problem that the first law appears to be a special case of the second, a possible solution is to narrow the scope of the second law until it no longer covers the first.

\cite{Koslow1969} suggests that the second law is only meant to cover the case of non-zero Forces. He suggests that Newton thought of forces as external causes and the second law is used when these causes are present. The first law is needed for the case these causes are absent:

“Newton believed that forces which act on a body are external causes whose effect is the body’s motion. . . . Newton’s second law, I suggest, does not state without qualification that the total force on a body equals mass times acceleration. This law has sense only if there are external causes acting upon a body. That is, the second law should be read as follows: if there are forces acting upon a body, then the total of all these forces equals the product of the body’s mass and acceleration . . . A statement of what happens when causes are present does not entail a statement describing what happens when those causes are absent.”~\citep[551-552]{Koslow1969}

Cohen and Whitman offer a different explanation, that in some sense narrows the scope of both laws. They claim that second law deals with impulse forces, whereas the first law, based on the examples Newton offers to it, refers to continually acting forces:

“Thus a possible clue to Newton’s thinking is found in examples used in the discussion of the first law, analyzed below, each of which is (unlike the forces in law 2) a continually acting force. Accordingly, we may conclude that law 1 is not a special case of law 2 since law 1 is concerned with a different kind of force.”~\citep[110]{newton1999principia}

Cohen and Whitman themselves acknowledge a significant weakness of this explanation, that Newton sometimes applies his second law for continually acting forces as well~\citep[112]{newton1999principia}.

\cite{Hoek2023} resolves the independence problem in yet a different way: he both augments the first law and restricts the scope of the second law, which is why we list this explanation here.

Hoek claims that the usual meaning attributed to the first law is mistaken, and it should be understood in a stronger sense. In the traditional meaning, which Hoek terms the “weak reading,” the law applies to bodies that no forces act on them, or in other words, it applies to free particles. The “strong reading” he offers is different, stating that any acceleration of a body is caused by a force:

“Every change in a body’s state of uniform rectilinear motion is compelled by the forces impressed on that body.” \citep[62]{Hoek2023}

Hoek justifies this strong reading by various means: analysis of the original text and the examples Newton describes for its use. One further such justification is demonstrating how the strong reading solves the independence problem. For this Hoek offers an interpretation of the second law as well. Instead of reading it as a relationship between the total force and the total acceleration, Hoek suggests it should be read as a relationship between a single force and the acceleration caused by it:

“Accordingly, the ‘change in motion’ that $L_{II}$ [the second law] describes is a ‘change in motion due to a particular impressed force’” ~\citep[65]{Hoek2023}

The extension to multiple forces is handled not by the second law, but by Corollary 2. With this interpretation, the second law does not rule out the possibility of acceleration due to other causes which are not forces. This is the gap the first law is intended to fill:

“But – and this is the crucial part – it is the First Law, not the Second, which tells us that any additional diversion from the state of motion would have to be due to further impressed forces.”~\citep[66]{Hoek2023}

The weak reading of the first law is sufficient to rule out the Aristotelian worldview. For example, it rules out that the natural tendency of an earthly object is towards the center of the Earth. But more generally, if a body is already subjected to forces, then the weak reading cannot rule out the possibility that this body has some additional acceleration on top of that caused by the forces. The strong reading eliminates this possibility altogether, explicitly stating every acceleration is caused by a force.

\section{Novel Explanations}
\label{sec:novel}
\subsection{The formal explanation}
\label{sec:formal}

In this section we propose a new solution to the independence problem, which relies on an understanding of the mathematical formalism used in the \emph{Principia}.

Newton chose to formulate the \emph{Principia} in terms of Euclidean geometry. 
Several motivations for this choice have been proposed~\citep{Densmore2003,Maudlin2012}. One of them is
 that, at the time, number theories did not rest on rigorous mathematical foundations, while Euclidean geometry did.
Regardless of the motivation, the fact of the matter is that the theorems presented in the \emph{Principia} are phrased using Euclidean concepts and their proofs rely on known geometric theorems, sometimes citing Euclid’s \emph{Elements} directly. 

Due to this choice of mathematical framework, the second law was not presented in the \emph{Principia} as the numerical formula $F=ma$, but stated verbally:

"A change in motion is proportional to the motive force impressed and takes place along the straight line in which that force is impressed."~\citep{newton1999principia}

Even though it is expressed in words, it is still a mathematical law, stating that two magnitudes, force and change in motion, are in proportion. We argue that it should be understood in Euclidean terms, as does the rest of the mathematics in the \emph{Principia}. Doing so makes it clear that the second law is not applicable to the cases of either no force or no acceleration, hence a separate law was added to cover this case.

Our first argument in support of this claim draws on Euclid’s definition of proportion. We quote here the three relevant definitions from Book V of Euclid’s \emph{Elements}:

\noindent “4.  (Those) magnitudes are said to have a ratio with respect to one another which, being multiplied, are capable of exceeding one another.\\
 5. Magnitudes are said to be in the same ratio, the first to the second, and the third to the fourth, when equal multiples of the first and the third either both exceed, are both equal to, or are both less than, equal multiples of the second and the fourth, respectively, being taken in corresponding order, according to any kind of multiplication whatever.\\
 6. And let magnitudes having the same ratio be called proportional.”~\citep{Fitzpatrick2008}

Definition 4 states that to have a ratio, magnitudes should be capable of exceeding one another when multiplied. This means that a zero magnitude cannot have a ratio with any other magnitude, zero or non-zero. Definition 5 specifies when the ratios of two pairs of magnitudes are said to be the same, and definition 6 states that in this case they are called proportional. Hence we can conclude that a zero magnitude is not in proportion with any other magnitude. 

Since Newton’s second law uses the Euclidean term \emph{proportion}, it is clear it applies only to non-zero force and acceleration. In contrast, in the algebraic formula $F=ma$, where the variables represent numbers or vectors, we can substitute $F=0$ and obtain the first law. We cannot do so in the Euclidean formulation of the second law, because the Euclidean definition of proportion becomes inapplicable. Thus, the case of a zero force needs a separate treatment, and this is the role of the first law: to provide this separate treatment.

One might object to the above conclusion by arguing that mathematics has evolved considerably since Euclid's time, and that although Newton used Euclidean geometry, he may not necessarily have adhered to every detail of Euclid's original definitions. We can test this by examining Newton's use of the term proportion in the \emph{Principia}.

One relevant passage appears in Section I of Book 1, which deals with the method of first and ultimate ratios. The Scholium to this section contains the following text:

"It may be objected that there is no such thing as an ultimate proportion of vanishing quantities, inasmuch as before vanishing the proportion is not ultimate, and after vanishing it does not exist at all."~\citep{newton1999principia}

This passage raises the possibility that no proportion exists in certain cases and identifies such a case explicitly: when the quantities have vanished. This is presented as a common-wisdom objection to the methods introduced in Section I. Newton responds to this objection by clarifying that the ultimate proportion is not the proportion after the quantities vanish, but the proportion determined by the ratios with which they vanish. This indicates that, for Newton---as well as according to the common understanding of his time---proportion was not defined in the case of vanished quantities.

We discuss Newton's use of proportion further in Section~\ref{sec:discuss_transition}, citing previous work that supports our claim.

A second argument in support of our view is the general approach in Euclidean geometry of distinguishing between a magnitude and the absence of a magnitude as two distinct cases. In Euclid’s \emph{Elements}, magnitudes can be of different kinds, such as lines or angles, but they are always taken to be non-zero~\citep[363]{Grattan-Guinness1996}. For example, the definition of an angle in Book I reads:

“8. And a plane angle is the inclination of the lines, when two lines in a plane meet one another, and are not laid down straight-on with respect to one another.”~\citep{Fitzpatrick2008}

The last part of this definition rules out a zero angle.\footnote{The definition does allow for curvilinear angles, such as the angle between a circle and its tangent, also known as a horn angle. Such angles are mentioned in the \emph{Elements}, though rarely (Book 3, proposition 16), and are still considered different than zero~\citep{Grattan-Guinness1996}.}
According to this definition, if we place two lines directly one on top of the other, then this is not a special case of an angle whose size is zero. It is a different case, where we can say that an angle is absent. 

There is evidence in the text of the \emph{Principia} that Newton maintained this separation between the case of non-zero magnitudes and the case of their absence. For example, the first law itself distinguishes between two states: “rest” and “moving uniformly.” It is clear from this distinction that “moving” here refers to a non-zero velocity. A similar separation appears in the explanation that follows the second law, where Newton describes how the new motion generated by a force is added to a body’s prior motion:

"And if the body was previously moving, the new motion [. . .] is added to the original motion [. . .]"~\citep{newton1999principia}

The condition that opens the quoted passage again shows that Newton takes care to distinguish between the cases: if the body was previously at rest, then it had no motion, and there is nothing to which the new motion can be added.

Similarly, we suggest that the two magnitudes to which Newton’s second law refers, change in motion and force, are non-zero. The case of their absence is treated separately by the first law, a separation that reflects the conventional practice of Euclidean geometry.

A third argument in favor of our approach is that it suggests that the reason for the separation between the laws is technical and, in some sense, trivial. This triviality is a strength of our view, since it answers another question: Why did Newton not address the independence problem himself? We know that he paid much attention in the \emph{Principia} to the logical foundations of his work. Most of the other reasons we have reviewed in this paper seem important and interesting enough to be communicated to the \emph{Principia}’s readers, and Newton would have explained them in the Scholium section. Silence supports the case Newton did not see here any special difficulty requiring explanation for 17\textsuperscript{th}-century readers.

In summary, the algebraic formula $F=ma$ covers the case $F=a=0$ as any other valid satisfying assignment. Describing this case in an additional, separate axiom is superfluous. But when the underlying mathematical language is unable to handle the case of zero together with the non-zero magnitudes, this separation into two axioms is the simplest and most natural choice.\footnote{It is noteworthy to compare these points with a related discussion by Aristotle~\citep[46]{Aristotle2000}. Aristotle starts from the assumption that the distance that a body endowed with weight moves naturally is proportional to its weight. He goes on to prove that a weightless body will not move at all. His argument is as follows: If the weightless body moves a certain distance AB, and a body endowed with weight moves some other distance AC at the same time, then we can use the proportion assumption to construct another body endowed with weight that moves the distance AB. Aristotle concludes that having both the weightless body and the body endowed with weight moving the same distance is absurd.  However, this absurdity stems from physical intuition, and this argument is compelling from the physical point of view. A similar argument can be adapted to Newton’s framework: The case of a force that generates the same acceleration as no force at all is physically absurd. Mathematically however, because proportionality does not deal with the case of the absence of one of the magnitudes, the case of no magnitude is mathematically undefined. And we could from the purely formal mathematical point of view associate any arbitrary acceleration with no force. Our argument is that this is precisely the conflict Newton wished to avoid by adding the first law. The proof from Aristotle’s book is followed by a second argument, that is not directly related to this paper, but still is interesting in relation to Newton’s work. In this segment Aristotle uses a similar style of argument to prove that there cannot be a body that is entirely weightless. Here and in other places we can see Aristotle had some anticipation of Newton’s concept of mass.}

\subsection{The logical explanation}
\label{sec:log}
We now propose a second novel explanation to justify stating the first law as a separate law. According to this explanation, even if the second law is formulated algebraically, in which case our previous explanation does not hold, then still the first law has a role to play. This role is to require a more fundamental natural property. It can continue to to hold even if it turns out the second law is different than the one described in the \emph{Principia}.

The first law states that in the absence of a force a body will preserve its state of motion. If a body is under the influence of forces then its velocity might change in direction or magnitude. But the first law does not specify how it will change, and allows for any relationship between force and acceleration. It is the second law that describes such a specific relationship: proportion.

Thus, we can view the first law as a more general principle. Imagining possible alternative worlds, there could be worlds where the first law holds, but there is a different quantitative relationship between force and acceleration, for example $F=\sqrt{ma}$ or $F=ma^2$. That is, each such alternative world has a different second law, all of them consistent with the same first law.

In the \emph{Principia}, the role of the second law is to specify this quantitative relationship. In its algebraic form, $F=ma$, the first law becomes redundant in the sense that it adds no information that is not already contained within the second law. But the importance of the first law, we argue, stems from the fact it can remain a valid axiom even if the more specific relationship is refuted. We note that this actually occurred: Relativity changed the second law, but the first law remained the same~\citep[17]{Taylor2005}. 

In this view, the fact that the first law can be obtained from the second by substituting $F=0$ shows that the quantitative relationship set down by the second law is compatible with the first, but it makes neither redundant.

\section{Discussion}
\label{sec:discussion}

\subsection{The transition from geometry to algebra}
\label{sec:discuss_transition}

Our novel formal explanation relies on the language of Euclidean geometry, and specifically on the definitions related to proportion in Euclid’s \emph{Elements}. A possible objection to our view, already mentioned in Section~\ref{sec:formal}, is that at the time of the publication of the \emph{Principia}, the language of mathematics was already undergoing significant changes that must be taken into account. In this section, we cite previous work on this subject that includes discussions of the arithmetization of proportion. These works provide additional support for our interpretation.

Between the 17th and 19th centuries, mathematics underwent a shift from a geometric to an algebraic mode of reasoning, a transition that profoundly affected both mathematical practice and its application in the sciences. 
\cite{goldstein2000matter}, reviewing changes in  published editions of Euclid's \emph{Elements} during this period, summarizes this transition as follows:

"In 1600, virtually all higher mathematics and quantitative science utilized geometric proof. In the nineteenth century, however, Carl Friedrich Gauss
termed arithmetic the queen of mathematics, and by 1900, analysis had supplanted geometry as the foundation for higher mathematics and the physical sciences."~\citep[36]{goldstein2000matter}.

Goldstein identifies two schools of editors of Euclid's \emph{Elements} active during this period. One group remained loyal to treating proportion as a relation between geometrical magnitudes and retained Euclid's original definitions. The second group recast geometry in algebraic terms, including the definitions of proportion. A drawback of these new approaches was the limitations inherent in the contemporary understanding of arithmetic regarding irrational numbers. The following passage summarizes how the two groups viewed these limitations:

"Continuing a tradition begun by geometers who had discovered incommensurable proportions that could not be expressed as ratios among whole numbers, traditionalist commentators such as Stone and Keill argued that numerical ratios pertained only to rational numbers as in Book VII of the Elements. Arithmetic and algebra could not deal with either incommensurables or irrational proportions; these could be expressed only as geometric magnitudes. More progressive editors did not dispute this distinction so much as they limited its
significance."~\citep[38]{goldstein2000matter}

Thus, in Newton's time, the geometrical definition of proportion was still the only one that could be universally applied in every case.

~\cite{sylla1984compounding} expands on the roots of these two views and links them to the \emph{Principia}. The first view belongs to a tradition rooted in the definitions of Book V of Euclid's \emph{Elements}, based on geometrical magnitudes. The second is rooted in a commentary by Theon of Alexandria on Ptolemy's \emph{Almagest}, which identifies ratios with numbers. 

Sylla finds evidence that the \emph{Principia} is loyal to the first tradition. When compounding ratios, Newton uses terms such as \emph{dupla} or \emph{dimidata} to refer to squares or roots of ratios. He used the same terms with integers, but there they refer to double or half. These differing meanings indicate Newton did not identify ratios with numbers. 

This terminology confused a contemporary mathematician, Clerke, who followed the second tradition. A correspondence between him and Newton led Newton to revise these terms in the  \emph{Principia}'s second edition. However, he did not do so fully and consistently. Sylla concludes:

"On the basis of this evidence and of other evidence to be found in Newton’s mathematical works, I believe that although Newton changed the terminology of the Principia to answer Clerke’s complaints, he never did give up the view that ratios are different from numbers. Despite the rapid development of analytic geometry in his day and despite his own significant contributions to it, Newton was not a mathematician who easily identified the entities and operations of the different branches of mathematics. Rather, he attempted to preserve or restore the purity or rigor of each branch of mathematics and had plans to write a work of geometry using only strictly geometrical methods."~\citep[15]{sylla1984compounding}

~\cite{grosholz1987some} provides further insight into Newton's use of proportion. She describes the first tradition as based on Definition 4 that we quoted in our formal explanation:

"The leading principle of this theory of ratio is the Eudoxian or Archimedian axiom, which states: 'magnitudes are said to have a ratio to one another which are capable, when multiplied, of exceeding one another.'"~\citep[210]{grosholz1987some}

Due to this definition, only magnitudes of the same kind can have a ratio. For example, an angle and a line segment cannot have a ratio. However, proportion can involve magnitudes of different kinds. For example, two angles and two line segments can participate in a proportion if their ratios satisfy Definition~5. 

The second tradition, in contrast, homogenizes magnitudes occurring in ratios and proportion, treating them uniformly as numbers~\citep[212]{grosholz1987some}. Grosholz reiterates the weakness of this second tradition: that it restricted numbers to rational numbers only. But she identifies an even deeper shortcoming. She argues that the homogenization of magnitudes raises philosophical difficulties for the application of geometry to physics, due to the differing nature of the magnitudes involved: 

"More generally, philosophers of science cannot afford to forget certain fundamental questions which the second tradition suppresses. What justifies the claim that physical magnitudes
(like distance, or time) stand in relations to each other similar to relations between numbers? Or that geometrical magnitudes do likewise?"~\citep[212]{grosholz1987some}

Thus, for for the \emph{Principia}, the first tradition was better suited. Grosholz notes several examples of this, one of which arises in Proposition XI:

"Finally, to reiterate my earlier point, Newton’s expression of his result as 'the centripetal force will be inversely as the solid $\frac{SP^2\times QT^2}{QR}$' allows him to exploit the proportion-idiom of the first tradition to relate and yet discriminate heterogeneous physical magnitudes, lines and areas, and finite and infinitesimal magnitudes."~\citep[216]{grosholz1987some}

She further identifies adherence to the first tradition in Newton's way of compounding ratios, which follows the more restricted rules of this tradition. Yet she finds specific sections in the \emph{Principia} where a different notation for compounded ratios is used, which Grosholz interprets as evidence of occasional use of the second tradition. She claims these occur for geometrical magnitudes that do not represent physical magnitudes:

"The magnitudes on the right-hand side, [ . . . ] are all constant finite geometrical line lengths [ . . .] with no physical import. Having no reason to think of them otherwise than as numbers, Newton can handle them according
to the second tradition, multiplying them like quotients."~\citep[219]{grosholz1987some}

In conclusion, the analysis of the historical context cited above provides substantial support for our formal explanation. It strengthens the case that Newton represented physical magnitudes using geometrical ones and treated them in accordance with the original Euclidean definitions of proportion. These definitions, as we argue in Section~\ref{sec:formal}, preclude applying the second law to zero force and zero acceleration.

\subsection{Assessment of the plausibility of different answers}

In Section~\ref{sec:review} we have reviewed a range of solutions to the independence problem approaching this question in different ways, and in Section~\ref{sec:novel} we offered our novel explanations. In this section we discuss the plausibility of these various accounts. 

Let us begin with our main novel answer, the formal explanation. In our view, it has been firmly established. We have shown that the definitions of Euclidean geometry restrict the second law to non-zero magnitudes. Furthermore, we have provided evidence of the \emph{Principia}'s commitment to these definitions and to the broader Euclidean convention of treating zero magnitudes as a separate case: a passage indicating that no proportion exists between zero magnitudes, together with several examples of the separate treatment of zero and non-zero magnitudes. Lastly, we have cited additional work claiming that the \emph{Principia} remained faithful to the Euclidean conception of proportion.

Hence, the puzzle of the first law's separate statement is resolved in a simple and straightforward way, and no further reasons are required. Nevertheless, it remains possible that Newton had additional motivations. Let us now discuss the plausibility of such additional reasons.

The first category of solutions we reviewed is the "Conceptual precedence" category, claiming that the second law cannot be fully understood without the first law. We think, however, that this has no support by the \emph{Principia}'s text. The "Axioms, or the Laws of Motion" section is preceded by a "Definitions" section, clarifying concepts such as quantity of motion, force, and absolute space. Thus, it is unlikely that the laws are meant to define any of these concepts. More generally, this structure of the text seems to indicate the conceptual framework is intended to be fully established before the laws are introduced.

One explanation in this category does require some additional attention: the suggestion of Pourciau and others that the role of the first law is to define what constitutes natural motion, while the role of the second law is to define how forces change this motion. We think this explanation, too, mismatches the text. First, the Definitions section already describes natural motion, both in the definition of vis insita (Definition 3) and in the definition of an impressed force (Definition 4). 

Second, and perhaps more importantly, the second law uses the term  'motion' as characterized by both magnitude and direction. This is evident, for example, in the explanation that follows it (we already quoted part of this passage in Section~\ref{sec:formal}), which repeatedly refers to both the direction and the magnitude of motion:

"And if the body was previously moving, the new motion (since motion is always in the same direction as the generative force) is added to the original motion if that motion was in the same direction, or subtracted from the original motion if it was in the opposite direction or, if it was in an oblique direction, . . ."~\citep{newton1999principia}

 Characterized by both direction and magnitude, it is clear what unaltered motion is: a motion with a constant magnitude and a constant direction. This means steady movement in a straight line in a Euclidean space (Euclidean space as a description of space is obvious in the 17\textsuperscript{th} century as no other alternatives were conceived). Thus, when the second law states that a change in motion is proportional to the force we can deduce that this change occurs with respect to unchanged motion, which is steady movement in a straight line. And therefore, the first law is not needed as preqrequisite to the second.

A second categeory we reviewed included explanations that accept that the first law is redundant. One such explanation is historical influence. There is indeed clear textual evidence for some historical influence.
However, Newton's laws appear under the title "Axioms, or the Laws of Motion". The use of the word “Axioms” is a strong indication that Newton did think of all three laws as independent. Thus, we think that it is unlikely Newton thought of the first law as redundant.

Some additional solutions we included in this category claim that while the first is law is redundant in the \emph{Principia}, it describes an important special case, so important that it warranted explicit mention. Our second novel explanation, the one we termed the logical explanation, belongs to this type. It derives the importance of the first law from it being more general and fundamental. Other authors we reviewed derived the first law's importance from other reasons. We think these views are reasonable, though they remain speculative in the absence of further supporting evidence.

The third and last category was solutions to the independence problem that narrow the scope of the second law. Our novel formal explanation belongs to this group as well. Another solution in this category
is due to Koslow, who similarly suggests that the second law is meant to cover only the case of non-zero forces. According to Koslow, the second law states what happens when causes are present, and this does not entail what happens when causes are absent. However, we argue that Koslow's explanation does not uncover the root cause of the separation between the two laws. The second law is stated as a mathematical law, and modern mathematical tools quantify absence using zero. Thus, there is no logical constraint prohibiting such laws from covering both the case that causes are present as well as the case that they are absent, as does for example the formula $F=ma$. The true root cause, we argue therefore, lies in the specific mathematical language used by the Principia, which does prohibit treating magnitudes and their absence as a single case.

Another solution in this category is Hoek's, which changes the traditional reading of both the first and second laws. This explanation is motivated by several reasons, and not solely by the independence problem. Nevertheless, in our view, it requires interpreting the first law in a way that may be difficult to reconcile with the law’s literal wording.

\section{Conclusion}
\label{sec:conclusion}

The main contribution of this paper is to offer a novel solution to the independence problem: the formal explanation. In our opinion, it is a strong, well established solution. We further offered a second possible explanation, the logical explanation, and provided a comprehensive review of previously suggested answers.

The question of the independence of Newton’s first law continues to surface in modern physics classrooms, as well as in textbooks and popular science discussions. We hope that our paper contributes to a more thorough, historically accurate and philosophically plausible account of the issue.

\backmatter

\section*{Declarations}


\bmhead{Funding}
No funding was received to assist with the preparation of this manuscript.


\bmhead{Author contributions}
I.Y. conceived of the novel approaches. I.Y. and E.A. jointly developed them and wrote the manuscript.


\bibliography{sn-bibliography}

\end{document}